# The Informational Content in Lepto-Variance and Its Relation to Higher Moments

**Vassilis Polimenis**

University of Limassol, Limassol, Cyprus
Email: polimenis@yahoo.com

## Abstract

Lepto-regression is defined as the machine learning process of constructing a Regression Tree of a target feature on itself. It is a novel, model-free method potentially revealing information on important sample structure properties. But it is yet not clear what the informational content of lepto-variance is and how it is related to other well-known statistics of a sample. One significant finding is that 58% of the historical US stock return variability is 1-bit lepto-variance that can not be explained by any financial factor. The central question investigated in this paper is to use small normal N(0, 1) drawn samples to explore how the 1-bit sample lepto-variance and lepto-ratio relate to sample variance, skewness and excess kurtosis. Using a large sample simulation, the lepto ratio of a normal is found to converge to 36.3%. For smaller normally distributed simulated N(0, 1) samples, while lepto-variance itself is highly correlated to sample variance, lepto-variance as a fraction of total variance is highly correlated to excess kurtosis. Both lepto-variance and lepto-ratio are orthogonal to sample skew. Another finding is that while lepto-ratio is strongly correlated to lepto-variance it remains orthogonal to sample variance.

## 1. Introduction and Main Findings*

The ***lepto-regression*** technique, defined in Polimenis (2022) and (2024), as constructing a Regression Tree (RT) of a target feature on itself, is a novel, model-free method potentially revealing important sample structure properties. Based on the principle that sorted splits minimize MSE, lepto-regression provides an upper

*Some of the results in this paper were presented in the 10th Indonesian Finance Association International Conference, Oct. 2024 and the 23rd Annual Conference of the Hellenic Finance and Accounting Association, Dec. 2024.

bound on the variability explainable by an RT. The ***k-bit lepto-variance*** ($\lambda k^2$) is the portion of variance remaining after lepto-regression up to k steps, representing structure unexplainable by any set of features. The ***lepto-ratio*** is defined as the lepto-variance expressed as a fraction of total sample variability. Conversely, macro-variance at depth k ($\mu k^2$) is the maximum variance explainable by any feature combination. The empirical analysis on 96 years of daily US stock market returns shows that the 1-bit macro-variance accounts for 42% of total variability, leaving 58% as structure unexplainable by any 1-bit RT. The 2-bit lepto-variance reduces the unexplained portion to 26.3% of the total variance, with specific percentages of the initial 1-bit lepto-variance remaining within the left (42%) and right (47%) subtrees.

Initially this paper explores the lepto-variance of a uniformly distributed sample and provides a connection to well-known results from quantization theory. Quantization is a fundamental signal processing technique in modulation and analog-to-digital conversion of a signal that goes back as early as 1948. Then we proceed to the central question investigated in this paper; to find how the 1-bit sample lepto-variance and lepto-ratio introduced in Polimenis (2022) and (2024) relate to sample variance and other major distributional statistics and in particular, the 3rd and 4th cumulants of a sample, skew and excess-kurtosis, and also how lepto-ratio is related to the lepto-variance itself. To explore these relations, small normal N(0, 1) drawn samples are used (n = 100). The use of small samples facilitates the study of the relation of the various sample characteristics, as these moments exhibit significant variation in small samples. Finally, using a simulation approach, the lepto ratio of a large sample drawn from a normal distribution is shown to converge to 36.3%.

For smaller normally distributed simulated N(0, 1) samples, the main findings on lepto-variance are that lepto-variance is highly correlated to sample variance (var explains 16 bp out of a total lepto-variance variability of 36 bp). Average lepto-variance is 0.35 and is quite volatile with std = 0.06, min value at 0.174 and max value at 0.694. Samples with small variance have average lepto-variance 0.31, while samples with large variance have average lepto-variance 0.39. Finally, for small normal samples, 1-bit of sample variance explains 16 bp out of a total lepto-variance variability of 36 bp.

For smaller normally distributed N(0, 1) samples, the lepto-ratio is highly correlated to the 4th cumulant and excess kurtosis. The average lepto-ratio is 35% but is much less volatile than lepto-variance, with std = 3.33%, min value at 23% and max around 49%. Interestingly, the sample lepto-ratio is also correlated to lepto-variance itself. Lepto variance ratio is maximally correlated with excess kurtosis (0.78) and the 4th cumulant (0.75). Notably, for normal samples, both lepto-variance and lepto-ratio are orthogonal to sample skew. For every extra unit of kurtosis, the lepto-ratio percentage is expected to grow by another 5.25%.

## 2. Motivation

Both fields of quantitative finance and economics heavily rely on an understand-ing and modeling of return variability and risk. Regression models are foundational tools in this endeavor, when used to explain a continuous target variable based on various features. A key measure of a regression model's explanatory power is the coefficient of determination, or R-squared ($R^2$), which quantifies the proportion of the dependent variable variability explained by independent variables included in the model. It serves as a measure of how effectively independent variables may capture variations in the dependent variable.

In a linear regression setting, adding relevant independent variables can potentially increase the explanatory power of the model. However, careful consideration is needed in this process to avoid a phenomenon known as overfitting; i.e. when a model becomes too closely tailored to the specific data used for parameter estimation, capturing not only true underlying patterns but also noise present in a particular dataset. This can lead to models that seemingly perform well on the training/estimation part of the data but poorly on new, unseen data.

Within the realm of financial analysis, regressions on various market-wide factors are frequently employed to model stock returns.[1] For instance, company size (SMB), or book-to-market ratio (HML) factors are frequently used in models that regress stock returns on factors. Residual variance in such financial regressions—i.e. the portion of variance not explained by the market wide factors—is dependent on the specific factors chosen for the regression. Generally, incorporating additional financial factors into models can lead to reduction of such residual variance.

Regression trees (RTs) represent an alternative (to linear regression) machine learning approach that is commonly used in explaining a continuous target variable by recursively partitioning the sample input space. RTs are similar as a concept to decision trees but are adapted for regression tasks. They build a hierarchical structure where internal tree nodes represent split decisions based on a specific feature leading to further branching down based on the feature values relative to a split point. This process continues by recursively partitioning the sample on the basis of specific sample features and split thresholds, aiming to minimize residual target variance within each resulting subset. A constant value, typically the average of the subtree instances for that terminal node, is assigned to the instances reaching that leaf node. RTs provide an inherent binary split of the sample space, utilizing the residual sum of squares (sum of squared error) as the minimization criterion. The prediction for a novel unseen instance is the average value within the subset that it falls into based on its associated feature values, as these average values minimize the residual mean square error (MSE), provided that the new instance has been drawn from roughly the same distribution of values as the training sample.

RTs offer significant advantages, including easy interpretability and a simple visual representation, both of which render the decision-making process easy for humans to understand. Furthermore, RTs are non-parametric constructs capable of capturing complex, nonlinear relationships within the data and are robust to outliers. However, similar to the linear regression methodology, RTs also face the challenge of overfitting the training data, thus capturing noise rather than just underlying patterns in the data. Various regularization techniques such as pruning have been developed and are employed to help address the overfit issue and thus improve model generalization ability. Improving our understanding of the inherent explanatory power of an RT is therefore valuable because RTs serve as fundamental building blocks for more advanced ensemble methods. Ensemble methods like random forests and gradient-boosted trees combine multiple individual trees to enhance predictive performance and robustness.

[1]For example, Fama & French (1993).

Beyond the mechanics of model fitting and explanation, there is a significant motivation for analyzing variability that stems directly from the field of financial risk management. Financial risk management is a vast area of both academic study and practical application within banking and finance. The fundamental prerequisite for managing risk effectively is its proper quantification. This involves measuring volatility and correlations across the entire universe of investable assets. Understanding the underlying sources contributing to this volatility is of paramount importance.

Investors are particularly interested in identifying and understanding the factors that influence investment return volatility. This understanding is crucial because volatility directly impacts both the assessment of risk associated with an investment and the overall decision-making process regarding portfolio construction and asset allocation.

The academic and financial practitioner communities have long sought the introduction of model-free methods for analyzing return variability. Such methods would offer insights into variability without relying on potentially restrictive assumptions inherent in specific models. An example often cited is the volatility index (VIX), introduced by Whaley (1993). The VIX is sometimes referred to as the market "barometer" and is tradable on the CBOE. While a later calculation method developed by Demeterfi et al. (1999) aimed to be more model-free, the VIX calculation itself is still described as neither simple nor intuitive. This highlights the ongoing interest and need for simpler, more intuitive model-free approaches to volatility analysis.

Within financial analysis utilizing machine learning, financial factors are used as features with the goal of finding those that can explain a large fraction of the total stock return variance. The portion of stock return variance that remains unexplained by these broad market financial factors is categorized as idiosyncratic risk for that particular stock. The total risk of an investment is understood as the sum of two components: risks determined by exposure to market factors (market risks) and idiosyncratic volatility. Idiosyncratic volatility specifically represents the risk component unique to an asset that is not determined by or related (i.e., orthogonal) to broader market movements. Despite its importance, the implications of idiosyncratic volatility for asset pricing are still not fully understood, as indicated by various studies.

Ultimately, quantifying variance is critical for investors whose goals include diversifying portfolios and mitigating risk exposure. A novel statistical method that is model-free and simple could provide valuable assistance in analyzing total return variability. Such a method could empower investors to make more informed decisions, enhance their risk assessment capabilities, and construct portfolios that are better aligned with their tolerance for total risk and their investment objectives.

## 3. The Historical Lepto-Variance of US Stock Returns

Polimenis (2024) introduces and explores concepts related to analyzing sample variance using regression trees (RTs), particularly with a focus on US stock returns. The paper defines lepto-regression as the novel technique involving the construction of an RT where the target feature is regressed on itself. This simple and model-free approach offers the potential to reveal important properties about the structure of a sample dataset.

The basis for lepto-regression lies in how RTs work. Regression trees are constructed by recursively partitioning a sample based on chosen features and split thresholds, aiming to minimize the residual target variance within the resulting subsets. At each internal node, an RT sorts samples based on a selected feature and identifies a splitting point that maximizes the drop in variance from the parent node to its children.

The key observation supporting lepto-regression is that the optimal factor for reducing mean squared error (MSE) is the target variable itself. This is because, in terms of minimizing MSE, when splitting a sample, it is always beneficial to utilize a sorted split. A binary split of a sample into left (L) and right (R) subsets is defined as sorted if all target values in L are smaller than all target values in R. A lemma derived from Fisher (1958) and shown in Polimenis (2022) confirms that a sorted split is always beneficial for minimizing MSE in an RT, and therefore, the target variable itself is the best factor to use in terms of MSE drop. Using the target as a factor allows all sorted splits to be evaluated, thus providing an upper bound on the explained variance or, equivalently, a lower bound on the residual MSE. This is also related to Breiman et al. (1984) for decision trees with binary (2-class) target Y (see the discussion in Ripley, 1996 and Hastie et al., 2009).

Building on this, lepto-variance is defined as the residual MSE resulting from this lepto-regression process. It represents the portion of variance that cannot be mitigated by any regression tree. This concept is valuable because it serves as a measure of the inherent variance within a dataset at a specific tree depth. By establishing an upper boundary on the "resolving power" of RTs for a sample, lepto-variance provides insights into the intrinsic structure of the dataset.

The article introduces the concept that, at each RT depth level, the overall variance within a dataset is broken down into lepto-variance and macro-variance. The sample k-bit macro-variance is defined as the maximal variance that can be accounted for by RTs with depths up to k. Conversely, the k-bit lepto-variance ($\lambda k^2$) is the residual structure after the sample has been lepto-regressed up to k times, representing the variance that cannot be explained by any set of features. The total variance ($\sigma^2$) can thus be decomposed as $\sigma^2 = \mu k^2 + \lambda k^2$, where $\mu k^2$ is the k-bit macro-variance. The macro-variance can also be understood as the upper bound of sample variability that can be explained. The concept of decomposing variance in this manner is related to the 1-dimensional clustering problem and techniques used in cartography like the Jenks natural breaks classification method (Jenks & Caspall, 1971) for choropleth maps in cartography, one-dimensional clustering and to signal quantization.

To illustrate the concept, the paper provides simple examples of lepto-variance calculations for small equiprobable sets. For a {−0.5, 0.5} set with total variance 0.25, the only split separates the two members, resulting in zero residual (lepto) variance ($\lambda 1^2 = 0$) and macro-variance equal to the total variance ($\mu 1^2 = 0.25$). For {−1, 0, 1} (total variance $\sigma^2 = 2/3$), the optimal split is {−1} and {0, 1}, yielding $\mu 1^2 = 1/2$ and $\lambda 1^2 = 1/6$. For {−1.5, −0.5, 0.5, 1.5} (total variance 1.25), the optimal split is the balanced one into {−1.5, −0.5} and {0.5, 1.5}, centered at −1 and 1, respectively, with residual variance $\lambda 1^2 = 0.25$ and variance drop $\mu 1^2 = 1$. The lepto ratio ($lR^2$) is defined as the ratio of lepto-variance to total sample variance at a specific depth. For the 4-member example, $lR1^2 = 0.25/1.25 = 20\%$. Examples with a 6-member set {−1, 0, 1, 2, 3, 4} demonstrate that the optimal split is not always balanced and depends on the sample structure. The examples show how changing within-cluster variance affects the optimal split and the resulting lepto-variance ratio.

The empirical analysis section focuses on estimating the historical lepto-variance of US stock returns using historical daily market return data for a 96-year period from July 1, 1926, to June 30, 2022. The dataset comprises 25,272 daily returns. The total variance for the entire sample is approximately 1.167 (in percentage squared).

The 1-bit lepto-regression analysis for the US stock return vector (Mkt) shows that the optimal 1-bit split is a 30 - 70 balance. This split occurs for Mkt returns less than or equal to −0.264%. The two children subsets resulting from this split are centered roughly at −1% (for the smaller 30% subset) and 0.5% (for the larger 70% subset).

The 1-bit lepto-variance ($\lambda 1^2$) for this historical US stock return data is calculated as 0.678. Using the total variance of 1.167, the 1-bit macro-variance ($\mu 1^2$) is calculated as the total variance minus the lepto-variance: 1.167 - 0.678 = 0.489. This 1-bit macro-variance (0.489) equals almost 42% of the total US stock variability (0.489/1.167 ≈ 0.419). This implies a 1-bit lepto-ratio ($lR1^2$) of approximately 58% (0.678/1.167 ≈ 0.581), representing the structure in the data that cannot be removed (explained) by any 1-bit RT.

To provide context, Polimenis (2024) compares this lepto-regression analysis with a 1-bit RT analysis where US stock returns are regressed on the two Fama-French factors, SMB (size) and HML (value). The Fama-French three-factor model is a well-known asset pricing model based on a linear regression of excess returns on the market excess return, SMB, and HML. When using the entire historical sample, the paper finds that HML is more efficient than SMB and is thus chosen for the optimal 1-bit RT regression on these factors. This RT on HML is described as highly skewed and capable of explaining very little of the total historical US stock variability. The residual squared error from this regression is 1.1315, which is roughly 97% of the total MSE (1.167).

The paper introduces a statistic, $mR1^2$, representing the percentage of the sample macro-variance that a specific feature can capture with a 1-bit RT. Using a 1-bit RT, the Fama-French factors (SMB and HML) can only explain a small fraction of the total explainable MSE (sample macro-variance, $\mu 1^2 = 0.489$). HML explains 0.0355 of the total MSE, which translates to $mR1^2 = 7.26\%$ of the 1-bit macro-variance (0.0355/0.489). SMB explains 0.025 of the total MSE, corresponding to $mR1^2 = 5.11\%$ of the 1-bit macro-variance (0.025/0.489). Overall, HML slightly dominates SMB in terms of explained variance in this context.

The concept of lepto-variance can be extended to trees with a maximum depth greater than 1. As the depth of an RT increases, the residual variance generally decreases. The principle that the best split at any node is achieved by sorting the target itself remains locally valid, although the paper notes that in rare, degenerate cases (particularly with small sample sizes), the greedy nature of RTs might lead to a split that is locally optimal but suboptimal for capturing variation at deeper levels. However, for large sample sizes and relatively low depths (less than 3 - 4 splits), the distinction between average and maximum depth is unlikely to matter. The j-bit lepto-variance ($\lambda j^2$) is defined as the minimum residual MSE for an average depth j, achieved when the target is lepto-regressed on itself j times.

Polimenis (2024) also performs a 2-bit lepto-regression analysis for historical US stock returns. This involves further splitting the two subsets created by the initial 1-bit split. For the left subtree (returns ≤ −0.264%, 30% of samples), the optimal 2-bit split point is for returns larger than −1.884%.

The leftmost child (Mkt ≤ −1.884%) comprises the smallest 3.45% of total market returns. This subset has an average return of −3% and is highly volatile, with a residual MSE of 1.968.

The centermost part of the left child (−1.884% < Mkt ≤ −0.264%) comprises 26.5% of the total sample. This subset is substantially less volatile, with a residual MSE of 0.167. Out of the total MSE of 0.877 reaching the left subtree, 0.373 is identified as lepto-structure beyond the resolving power of the 2-bit RT within that subtree. This means 42% of the total variability of the left subtree is lepto.

For the right subtree (returns > −0.264%, 70% of samples), the optimal 2-bit split point is for returns larger than 1.145%.

The rightmost child (Mkt > 1.145%) comprises the highest 8.6% of the entire daily return sample. This subset has an average return of 2% and is highly volatile, with a residual MSE of 1.393.

The centermost part of the right child (−0.264% < Mkt ≤ 1.145) comprises the largest subsection, 61.5% of the total sample. This subset is substantially less volatile, with a residual MSE of 0.124. For the right subsample, the analysis shows that 47% of the total variance is lepto.

The overall 2-bit lepto-variance ($\lambda 2^2$) for the historical US stock return sample (total variance 1.167) is calculated by weighting the residual MSEs of the four resulting leaf nodes by their sample proportions: $\lambda 2^2 = (0.034 * 1.968) + (0.265 * 0.167) + (0.615 * 0.124) + (0.086 * 1.393) \approx 0.307$. This 2-bit lepto-variance (0.307) implies that the 2-bit lepto-variance equals 26.3% of the total variance (0.307/1.167 ≈ 0.263).

## 4. The Lepto-Variance of the Uniform Distribution

**Lemma.** The lepto ratio for the continuous uniform distribution is $lR^2 = 25\%$.

**Proof.** The continuous uniform distribution is an easy case to calculate the lepto-variance analytically, because the optimal split will be in the middle point $(a + b)/2$. To see this, observe that the cluster distance for any split is always $\frac{b-a}{2}$ regardless of the split point. Thus, the explained variance for a split at the qth percentile equals $q(1-q)\left(\frac{b-a}{2}\right)^2$ which maximizes for $q = 1/2$. This produces two children that are also continuous uniform distributions from a to $(a + b)/2$ and from $(a + b)/2$ to $b$.

The variance of the continuous uniform from $a$ to $b$ equals $\sigma^2 = \frac{(b-a)^2}{12}$. Given the optimal split in the middle, 1-bit lepto-variance equals the residual children variance

$$\lambda 1^2 = \sigma_l^2 = \sigma_r^2 = \frac{\left(b - \frac{a+b}{2}\right)^2}{12} = \frac{(b-a)^2}{48} = \frac{\sigma^2}{4}$$

Thus, the continuous uniform obtains a 1-bit lepto-variance ratio $lR1^2 = \lambda 1^2 / \sigma^2 = 25\%$ of the total variance. Notice that besides a low $lR1^2$ the continuous uniform is platykurtic with excess kurtosis $kyrt = -\frac{6}{5}$.

This level of lepto-ratio for the continuous uniform is related to the well-known result from quantization theory (Bennett 1948), that the *signal-to-noise ratio* (SNR) for uniform quantization (Pulse Code Modulation or PCM) is 6 db per bit (Oliver et al., 1948). To see why, take the

$$10\log_{10}\frac{\lambda 1^2}{var(y)} = 10\log_{10} lR1^2 = 10\log_{10} 1/4 \approx -6\text{ db}$$

The major difference between quantization theory and lepto-regression, is that in lepto-regression the support region of each tree leaf is determined by the sample (ex post) and not fixed a priori as in quantization. So in quantization, the goal is to produce an efficient encoding for any message, while in lepto-regression the goal is to study a specific sample.[2]

Similar to the continuous uniform distribution, we may find the lepto variance of its discrete analog for $n = 2^i$ elements. In the simplest case we have the equiprobable 2-member set {0, 1}. In this degenerate case, the only split is the separation of the two members that produces a drop of ¼ and the residual (lepto) variance will be zero.

The next case is the equiprobable 4-member set {0, 1, 2, 3}. In this case, the optimal split is balanced producing two equiprobable 2-member sets left and right, with residual variance equal to ¼. The two clusters are centered at 0.5 and 2.5, giving an inter-cluster distance of 2, and variance drop equals ¼ × distance squared i.e. drop = 1. This implies a total variance of ¼ + 1 = 1.25, and a lepto-ratio of 20% (i.e. −7 db).

We may continue in a similar fashion for the 8, 16, 32 etc. equiprobable uniform set of size $N = 2^i$ that is optimally split into two sets of size $2^{i-1}$ respectively with an inter-cluster distance of $N/2 = 2^{i-1}$ resulting in a variance drop of $\frac{1}{4} \times 2^{2i-2} = 2^{2i-4} = \frac{N^2}{16}$. From **Table 1** it is clear that as sample size increases, the lepto variance fast converges to ¼ of the total variance (and the limit SNR to approx. – 6). This is the limit continuous uniform case also as shown above. The Discrete Uniform distribution of size *N*, has known total variance equal to $\frac{N^2 - 1}{12}$.

**Table 1.** The lepto-variance for the discrete uniform with N points.

| *i* = log2(N) | sample size N | cluster distance | variance drop | lepto variance | total variance | Lepto ratio | Residual in db |
|---|---|---|---|---|---|---|---|
| 1 | 2 | 1 | 0.25 | 0 | 0.25 | 0.000000 | |
| 2 | 4 | 2 | 1 | 0.25 | 1.25 | 0.200000 | −6.9897 |
| 3 | 8 | 4 | 4 | 1.25 | 5.25 | 0.238095 | −6.2325 |
| 4 | 16 | 8 | 16 | 5.25 | 21.25 | 0.247059 | −6.0720 |
| 5 | 32 | 16 | 64 | 21.25 | 85.25 | 0.249267 | −6.0334 |
| 6 | 64 | 32 | 256 | 85.25 | 341.25 | 0.249817 | −6.0238 |
| 7 | 128 | 64 | 1024 | 341.25 | 1365.25 | 0.249954 | −6.0214 |
| 8 | 256 | 128 | 4096 | 1365.25 | 5461.25 | 0.249989 | −6.0208 |
| 9 | 512 | 256 | 16,384 | 5461.25 | 21845.25 | 0.249997 | −6.0206 |
| 10 | 1024 | 512 | 65,536 | 21845.25 | 87381.25 | 0.249999 | −6.0206 |

## 5. The Lepto-Variance of Small Normal Samples

In this section, the lepto-variance of samples drawn from a normal distribution is investigated. For a normal, skewness (skew) and excess kurtosis (kyrt) are zero. In a large normal sample all sample statistics will converge to their theoretical exact values (i.e. skew = 0 and kyrt = 0) and thus in large samples there is no variability that will allow us to study how they correlate to lepto-variance and lepto-ratio.

[2]For a review of Quantization theory see Gray & Neuhoff (1998).

Small samples drawn from a N(0, 1) distribution facilitate the study of the relation between lepto-variance for the sample and various other sample statistics, such as sample variance, skewness (skew), excess kurtosis (kyrt) etc.

For each of the 10,000 N(0, 1) drawn samples, Table 2 provides the sample variance (var), 3rd cumulant (cum3), 4th cumulant (cum4), skewness (skew), excess-kurtosis (kyrt) and the calculated 1-bit sample lepto-variance $\lambda 1^2$ and lepto-ratio $lR1^2 = \lambda 1^2 / \sigma^2$. For comparison to the quantization literature, the lepto-variance ratio is also presented as residual noise for the sample in decibels (db) via $10\log\left(lR1^2\right)$.

Table 2. (a) Lepto-variance and lepto-ratio of 10,000 simulated small (n = 100) normal N(0, 1) samples. (b). Correlation structure for a 10,000 small N(0, 1) samples (n = 100) simulation.

(a)

| n = 100 | var | cum3 | cum4 | skew | kyrt | lepto-var | lepto-ratio | db |
|---|---|---|---|---|---|---|---|---|
| count | 10,000 | 10,000 | 10,000 | 10,000 | 10,000 | 10,000 | 10,000 | 10,000 |
| mean | 0.991 | −0.002 | 0 | −0.002 | 0.001 | 0.35 | 35.345 | −4.536 |
| std | 0.142 | 0.249 | 0.519 | 0.245 | 0.497 | 0.06 | 3.336 | 0.412 |
| min | 0.518 | −1.05 | −1.949 | −1.003 | −1.191 | 0.174 | 23.052 | −6.373 |
| 25% | 0.893 | −0.155 | −0.325 | −0.161 | −0.344 | 0.308 | 33.041 | −4.809 |
| 50% | 0.984 | 0 | −0.073 | 0 | −0.077 | 0.347 | 35.28 | −4.525 |
| 75% | 1.082 | 0.151 | 0.241 | 0.156 | 0.249 | 0.388 | 37.571 | −4.251 |
| max | 1.648 | 1.356 | 6.213 | 1.047 | 4.417 | 0.694 | 48.921 | −3.105 |

(b)

| n = 100 | var | cum3 | cum4 | skew | kyrt | lepto-var | lepto-ratio | db |
|---|---|---|---|---|---|---|---|---|
| var | 1 | 0 | −0.012 | 0.002 | −0.007 | 0.831 | −0.006 | −0.005 |
| cum3 | 0 | 1 | −0.005 | 0.978 | −0.01 | 0 | −0.002 | −0.002 |
| cum4 | −0.012 | −0.005 | 1 | −0.009 | 0.96 | 0.421 | 0.75 | 0.74 |
| skew | 0.002 | 0.978 | −0.009 | 1 | −0.012 | 0 | −0.006 | −0.006 |
| kyrt | −0.007 | −0.01 | 0.96 | −0.012 | 1 | 0.425 | 0.782 | 0.772 |
| lepto-var | 0.831 | 0 | 0.421 | 0 | 0.425 | 1 | 0.546 | 0.545 |
| ratio | −0.006 | −0.002 | 0.75 | −0.006 | 0.782 | 0.546 | 1 | 0.998 |
| db | −0.005 | −0.002 | 0.74 | −0.006 | 0.772 | 0.545 | 0.998 | 1 |

In the simulation here, sample skew roughly varies from −1 to +1, and sample kyrt from roughly −1.2 to 4.4 (see Table 2(a)).

The 1-bit Lepto variance is around 0.35 and quite volatile with std = 0.06 and varies between 0.174 and 0.694.

The 1-bit Lepto variance ratio $lR1^2$ (sample lepto-variance as a fraction of total variance) is on the average 35.345% of total variance less volatile with std = 3.33% and may vary anywhere from 23% to 49% of the total sample variance.

## 6. Analysis of Lepto-Ratio

Lepto variance ratio $lR1^2$ is maximally correlated with excess kurtosis (0.78) and the 4th cumulant (Table 2(b)). Then, lepto ratio is also correlated to lepto variance itself. Notably, both lepto variance and lepto ratio are orthogonal to sample skew. It is not clear if such orthogonality will remain for samples generated by a skewed distribution. Another finding is that while lepto-ratio is strongly correlated to lepto-variance, and despite strong correlation of lepto-variance to sample variance, lepto-ratio remains orthogonal to sample variance.

Below we run a linear regression of the type

$$ratio = \alpha + \beta \cdot kyrt + e$$

The slope of the regression line is estimated by the correlation between $x$ and $y$ (0.782) and standard deviation of $x$ (0.497) and $y$ (3.336) via

$$b = 0.782\left(\frac{3.336}{0.497}\right) = 5.25$$

The explained R-squared of the regression is $R^2 = 0.782^2 = 61.2\%$.

For every extra unit of kurtosis, the lepto-ratio is expected to grow by another 5.25%.

Furthermore, given that excess kurtosis (kyrt) for the sample is centered at 0, the intercept equals the mean lepto-ratio value of 35.34 (Figure 1).

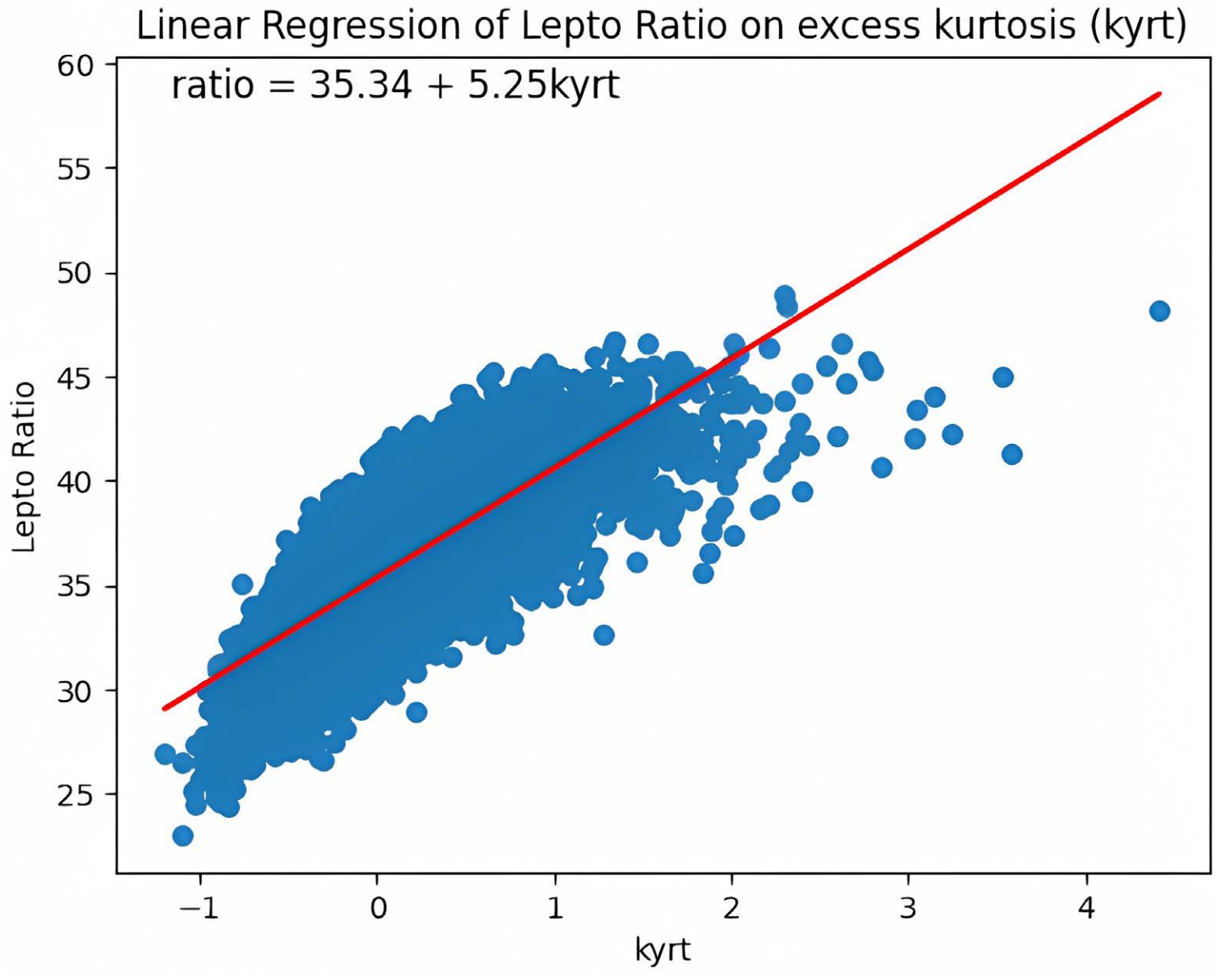

**Figure 1.** Linear regression of lepto ration on excess kurtosis (kyrt).

## 7. Regression Tree Analysis of Lepto-Ratio on Excess-Kurtosis

To get a better understanding of how lepto-ratio is explained, a regression tree based analysis is also presented below (Figure 2).

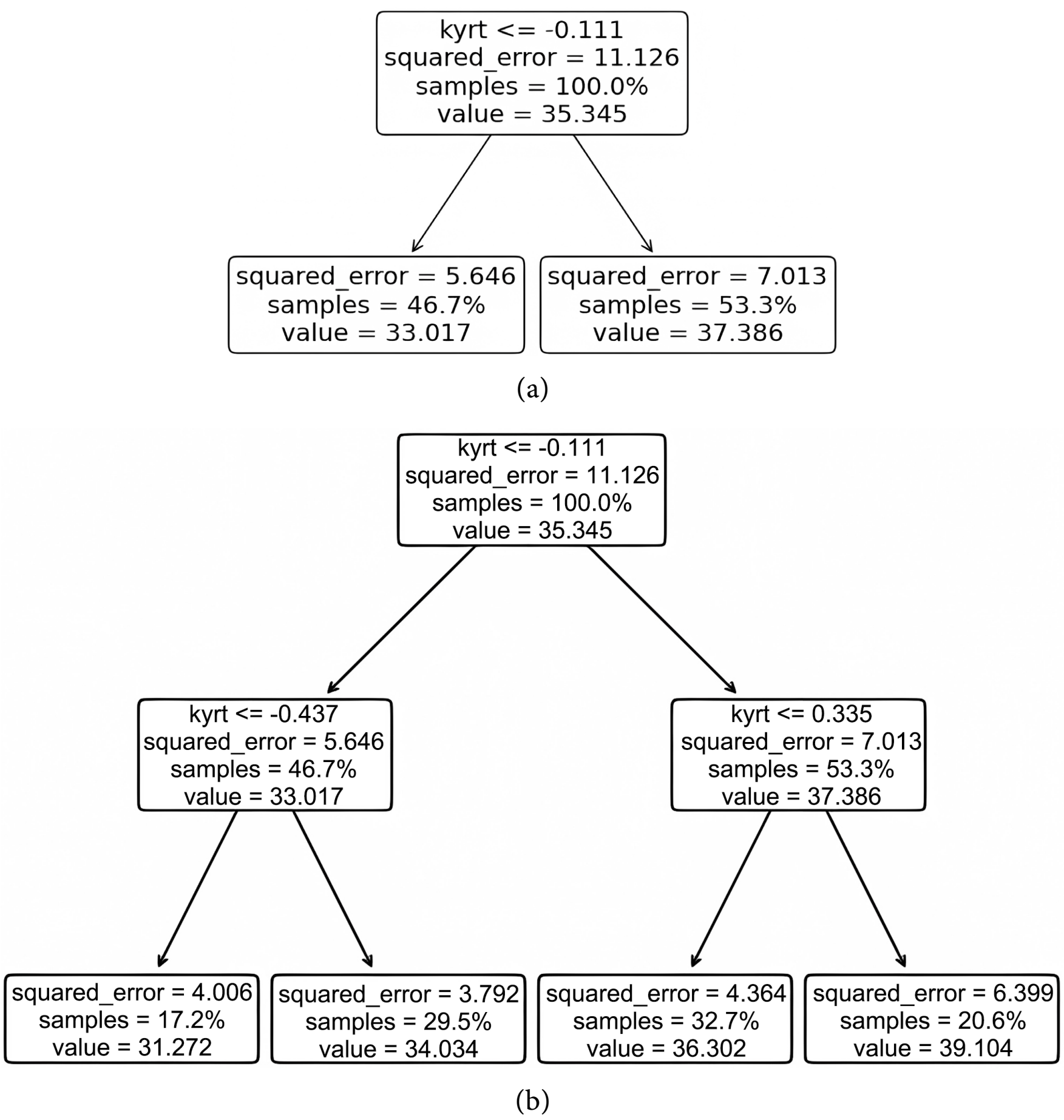

Figure 2. (a) Regression tree of *lepto-ratio on kyrt* - 10,000 N(0, 1) drawn samples. (b) Depth 2 Regression tree of *lepto-ratio on kyrt* - 10,000 N(0, 1) drawn samples.

Figure 2(a) presents the regression tree of *lepto-ratio* on var, cum3, cum4, skew, *kyrt* and lepto-variance itself. Excess kurtosis kyrt has the highest power in explaining lepto-ratio. Squared error is in basis points (1 bp = 1% × 1%) and average ratio value in percentage points. The 1-bit RT explains 4.6 bp out of a 11 bp total variability for a 6.4 bp residual MSE.

Figure 2(b) presents the 2-bit Regression tree of *lepto-ratio* on var, cum3, cum4, skew, kyrt and lepto. Again excess kurtosis kyrt has the highest power in explaining lepto-ratio even at the 2nd split.

For the 2-bit RT:

Residual MSE $= 0.172 \times 4 + 0.295 \times 3.792 + 0.327 \times 4.364 + 0.206 \times 6.4 = 4.55$

Thus, the 2-bit RT roughly explains 6.5 bp out of a 11 bp total variability for a 4.5 bp residual MSE.

## 8. Analysis of Lepto-Variance

Lepto-variance is highly correlated to sample variance (0.83). Secondarily, lepto-variance is also correlated with excess kurtosis at 0.425 [and the 4th cumulant].

A linear regression of lepto-variance (lepto-var) on sample variance (var) will have a least-square slope $b = \sigma_{xy} / \sigma_x^2 = \rho_{xy} \left( \sigma_y / \sigma_x \right)$ which equals $b = 0.831(0.06/0.142) = 0.351$. The regression line for the case of small N(0, 1) samples is given by $lepto = 0.0025 + 0.351\, var$ and shown in Figure 3.

For the full OLS regression statistics see the data below (Figure 4).

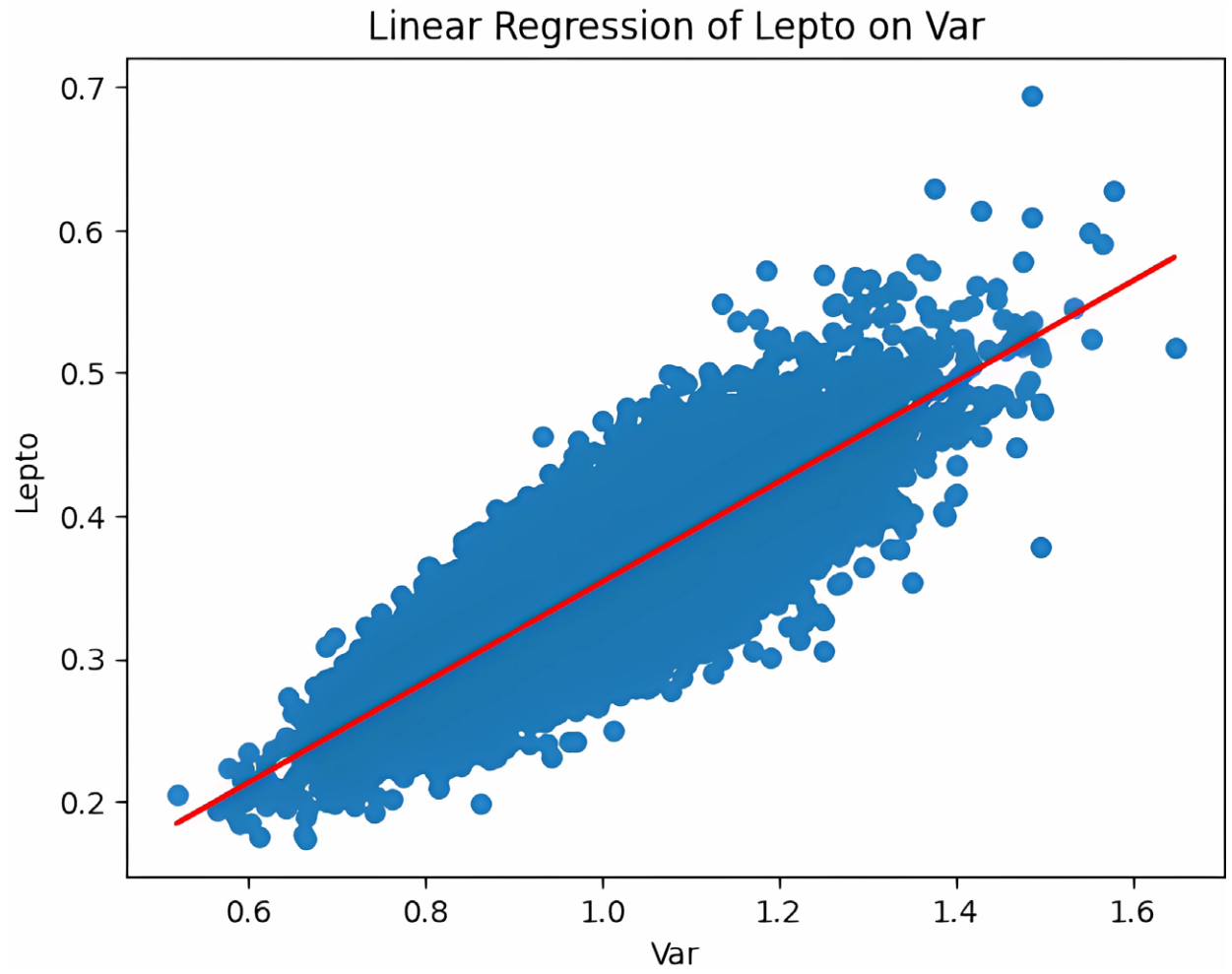

Figure 3. Linear regression of Lepto on var.

```
                            OLS Regression Results
==============================================================================
Dep. Variable:                  lepto   R-squared:                       0.691
Model:                            OLS   Adj. R-squared:                  0.691
Method:                 Least Squares   F-statistic:                 2.231e+04
Date:                Mon, 28 Oct 2024   Prob (F-statistic):               0.00
Time:                        06:25:10   Log-Likelihood:                 19824.
No. Observations:               10000   AIC:                        -3.964e+04
Df Residuals:                    9998   BIC:                        -3.963e+04
Df Model:                           1
Covariance Type:            nonrobust
==============================================================================
                 coef    std err          t      P>|t|      [0.025      0.975]
------------------------------------------------------------------------------
const          0.0025      0.002      1.053      0.292      -0.002       0.007
var            0.3509      0.002    149.372      0.000       0.346       0.356
==============================================================================
Omnibus:                       67.742   Durbin-Watson:                   2.016
Prob(Omnibus):                  0.000   Jarque-Bera (JB):               80.958
Skew:                           0.135   Prob(JB):                     2.63e-18
Kurtosis:                       3.348   Cond. No.                         14.0
==============================================================================
```

Figure 4. OLS regression results.

## 9. Regression Trees on Variance and Excess-Kurtosis

To get a better understanding of how lepto-variance is explained by sample var, a regression tree based analysis is also presented below (Figure 5).

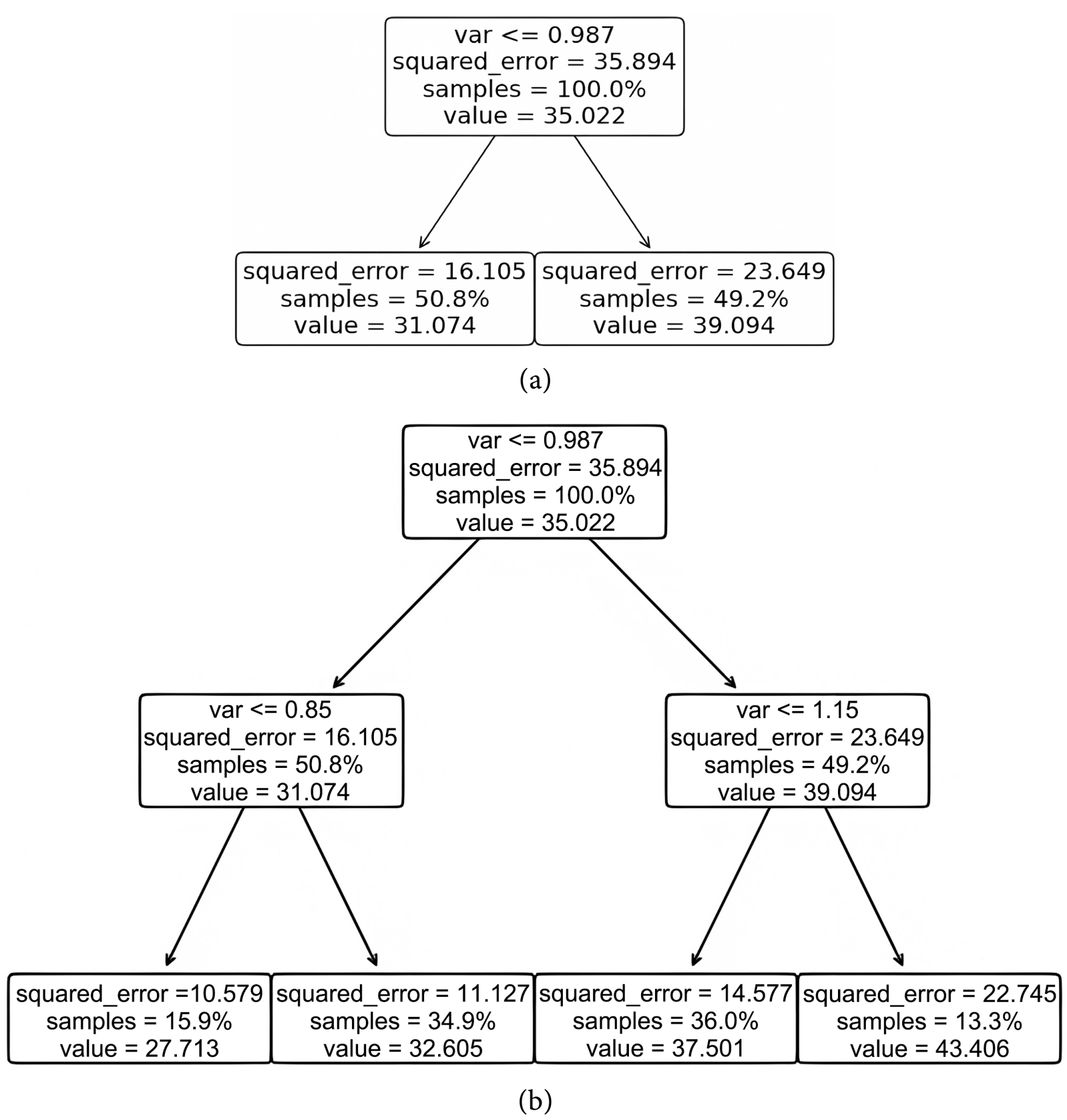

**Figure 5.** N(0, 1) drawn samples. 2-bit Regression tree of *lepto-variance* on var, cum3, cum4, skew, and kyrt. Sample variance *var* has the highest power in explaining lepto-variance for both the 1st and the 2nd split also. Optimal splits are balanced at the var = 0.85, 1 and 1.15 points respectively. This 2-bit RT produces a residual MSE of 14 bp. (a) Regression tree of lepto-variance on var - 10,000 N(0, 1) drawn samples. (b) Depth 2 Regression tree of lepto-variance on var 10,000 N(0, 1) drawn samples.

**Sample variance (var)** has the highest power in explaining lepto-variance. The mean lepto-variance for the simulation is 35%. Optimal 1-bit split criterion is based on sample var < 0.987 and results in a balanced split 51 - 49. Low variance samples have a mean lepto-variance around 31% while high variance samples have a mean lepto-variance around 39%. This 1-bit RT produces a residual MSE of 20 bp, thus var explains 16 bp out of a total lepto-variance variability of 36 bp (1 bp = 0.0001); i.e. 1-bit of sample var explains roughly $R1^2 = 45\%$ of the total lepto-variance variability.

To get a more clear picture, a simulation of large sample sizes is run and presented in Table 3.

To finally obtain the **convergence value** for the normal distribution a very large sample size 10 M is simulated for 10 times.

From Table 4, it becomes clear that the lepto-variance ratio of a normal variable is around 36.3%.

**Table 3.** Lepto-variance of 10 simulations of large sized (n = 100,000) N(0, 1) samples. Lepto ratio $IR1^2$ is on the average 36.3% (4.4 db) of total with std = 0.1%. $IR1^2$ is tightly spaced between 36.2% to 36.5% of total sample variance. The signal-to-noise ratio is between 4.377 and 4.408 db.

| n = 100,000 | var | cum3 | cum4 | skew | kyrt | Lepto var | Lepto ratio | db |
|---|---|---|---|---|---|---|---|---|
| **count** | 10 | 10 | 10 | 10 | 10 | 10 | 10 | 10 |
| **mean** | 0.999 | 0.003 | −0.001 | 0.003 | −0.001 | 0.363 | 0.363 | 4.396 |
| **std** | 0.004 | 0.007 | 0.012 | 0.007 | 0.012 | 0.001 | 0.001 | 0.01 |
| **min** | 0.992 | −0.005 | −0.025 | −0.005 | −0.025 | 0.362 | 0.362 | 4.377 |
| **25%** | 0.998 | −0.003 | −0.008 | −0.003 | −0.008 | 0.362 | 0.363 | 4.39 |
| **50%** | 0.999 | 0.001 | 0.003 | 0.001 | 0.003 | 0.363 | 0.363 | 4.4 |
| **75%** | 1.001 | 0.007 | 0.007 | 0.007 | 0.007 | 0.364 | 0.364 | 4.403 |
| **max** | 1.007 | 0.016 | 0.015 | 0.016 | 0.015 | 0.365 | 0.365 | 4.408 |

**Table 4.** Lepto-variance of 10 simulations of very large sized (n = 10,000,000) N(0, 1) samples. Lepto variance is always at 36.3% of total sample variance or 4.4 db.

| n = 10^7 | var | cum3 | cum4 | skew | kyrt | lepto var | lepto ratio | db |
|---|---|---|---|---|---|---|---|---|
| **count** | 10 | 10 | 10 | 10 | 10 | 10 | 10 | 10 |
| **mean** | 1 | 0 | 0 | 0 | 0 | 0.363 | 0.363 | 4.396 |
| **std** | 0 | 0.001 | 0.002 | 0.001 | 0.002 | 0 | 0 | 0.001 |
| **min** | 0.999 | −0.001 | −0.003 | −0.001 | −0.003 | 0.363 | 0.363 | 4.395 |
| **25%** | 1 | 0 | −0.001 | 0 | −0.001 | 0.363 | 0.363 | 4.396 |
| **50%** | 1 | 0 | −0.001 | 0 | −0.001 | 0.363 | 0.363 | 4.396 |
| **75%** | 1 | 0.001 | 0.001 | 0.001 | 0.001 | 0.363 | 0.363 | 4.397 |
| **max** | 1.001 | 0.001 | 0.003 | 0.001 | 0.003 | 0.364 | 0.363 | 4.399 |

## 10. Conclusion

The lepto-variance and lepto-regression are novel non-parametric statistical concepts introduced in Polimenis (2022) and (2024). They are related to similar methods developed in cartography, one-dimensional clustering and signal quantization. The central question investigated in this paper is to use small normal N(0, 1) drawn samples to explore how the 1-bit sample lepto-variance and lepto-ratio relate to sample variance, skew and kyrt. Then, using a large sample simulation, the lepto ratio of a normal is shown to converge to 36.3%. For smaller normally distributed simulated N(0, 1) samples, lepto-variance is highly correlated to sample variance and lepto-ratio is highly correlated to excess kurtosis. Both lepto-variance and lepto-ratio are orthogonal to sample skew. While lepto-ratio is strongly correlated to lepto-variance it remains orthogonal to sample variance. It is not clear if such orthogonality will remain for samples generated by a skewed distribution and further study is needed.

## Conflicts of Interest

The author declares no conflicts of interest regarding the publication of this paper.